\documentclass[12pt]{iopart}
\usepackage{amsmath}
\usepackage{amssymb}
\usepackage{graphicx}
\usepackage{dcolumn}
\usepackage{bm}
\usepackage[subnum]{cases}
\usepackage{soul}
\usepackage[utf8]{inputenc}
\usepackage{empheq}
\usepackage{color}
\usepackage{setspace}

\renewcommand{\v}[1]{{{\bf #1}}}
\renewcommand{\d}[1]{{\widetilde{\bf #1}}}
\newcommand{\dy}[1]{{\overset{\text{\tiny$\leftrightarrow$}}{\rm #1}}}

\newcommand{\rfig}[1]{Fig.~\ref{fig:#1}}
\usepackage{lineno}

\def\XXint#1#2#3{{\setbox0=\hbox{$#1{#2#3}{\int}$}
     \vcenter{\hbox{$#2#3$}}\kern-.5\wd0}}

\begin{document}


\title{Clausius-Mossotti Relation Revisited: Media with Electric and Magnetic Response}

\author{Lang Wang}

\address{Department of Electrical \& Computer Engineering, University of Illinois at Urbana-Champaign, Urbana, IL 61801, USA}

\author{Ilia L. Rasskazov}

\address{The Institute of Optics, University of Rochester, Rochester, NY 14627, USA}

\author{P. Scott Carney}
\address{The Institute of Optics, University of Rochester, Rochester, NY 14627, USA}
\ead{scott.carney@rochester.edu}
\vspace{10pt}

\begin{abstract}
A reexamination of the Clausius-Mossotti relation in which material with both electric and magnetic responses yields surprising results. 
Materials with indices near zero and with real parts less than zero, that is the real part of both the permeability and permittivity are negative, are found to emerge from the interaction of electric and magnetic responses in a self-consistent theory. 
The new results point the way to artificial and natural materials with exotic responses.
A simulation with $\sim 10^{10}$ particles shows good agreement with the analytical results.
\end{abstract}

\maketitle

\section{Introduction}

Effective medium theory (EMT) bridges the \textit{macroscopic} electromagnetic response (relative effective permittivity, $\varepsilon$, and relative permeability, $\mu$) of a composite electromagnetic medium with the \textit{microscopic} response (electric, $\alpha_{\rm e}$, and magnetic, $\alpha_{\rm m}$, polarizabilities) of its constituents~\cite{Sihvola1999,Choy2016,Markel2016MaxwellTutorial,Markel2016IntroductionTutorial}. 
The basic concept of the EMT is to characterize the response of a medium (for example, a suspension of plasmonic nanoparticles~\cite{Aspnes2011}, or aerosol particles~\cite{Mishchenko2016ApplicabilityParticles}) to light by the effective parameters.
The relation between microscopic electric response and macroscopic electric response was developed by Clausius, Mossotti~\cite{mossotti1850discussione,clausius1879mechanical}, Lorenz, and Lorentz ~\cite{lorenz1880ueber,lorentz1881ueber}, with a magnetic analogue proposed earlier by Poisson~\cite{poisson1825second}. 
While EMT has been put in a more general setting in the last 100 years, none has self-consistently addressed the interaction between the total electromagnetic field and the material that exhibits both electric and magnetic responses simultaneously.
The Clausius-Mossotti relation (CMR) and related EMTs~\cite{debye1912theorie,onsager1936electric,Lewin1947,garnett_xii_1904,garnett_vii_1906,bottcher1942theorie,kirkwood1936theory,McPhedran1997DynamicFormula,Sharma2000StudiesParticles} have not considered the electric field generated by the magnetic dipole moment induced on the particle and magnetic field by electric dipole moment.
EMT was first developed to explain the refractive index of natural materials which are weak- or non-magnetic.
In recent years, however, more materials have been discovered to have also strong magnetic responses, such as paramagnetic complexes/metalloproteins~\cite{MohadjerBeromi2019SynthesisReactions,Martel2014High-resolutionChemistry,Luchinat2012Solid-stateRestraints,Knight2012StructureNMR,Estephane2010AState,Flambard2009RevisitingO,Pell2019ParamagneticState,Vaidya2018SubstitutedPhenomenon,Adam2018CatalyticReactions}, magnetoelectric multiferroics~\cite{Gao2018StrongFluids,Dou2022Two-dimensionalSkyrmions,Vaz2010MagnetoelectricStructures,Spaldin2019AdvancesMultiferroics,Mundy2016AtomicallyMultiferroic,Nan1993,Nan2008,Jiang2007MagnetoelectricSintering}, metamaterials~\cite{Yang2022ObservationMetamaterial,Louis2018AModel,Dudek2019ImpactMetamaterials,Zhao2019IntegratingMetadevices,Duan2019BoostingMetamaterials,Koschny2004,Simovski2007,Tsukerman2011,slovick2014generalized,Zhang2015a,Krokhin2016,Tsukerman2017}, metasurfaces~\cite{Liu2021SimultaneousMetasurfaces,Cui2019LightMetasurface,Ignatyeva2020All-dielectricResonances,Abujetas2021Near-FieldModel,Zhang2022HuygensCoating,Brizi2022MagneticRegions},  nanoparticles~\cite{Schilder2017,Pellico2019Nanoparticle-BasedImaging,Xie2018Shape-Theranostics,Zhang2019ApplicationsCatalysis,Zhu2018MagneticApplications,Thorat2017EffectiveRelease}, periodic composites~\cite{Markel2012},  and mesocrystals~\cite{Qi2019SynthesisActivity,Du2020FeTherapy,Guo20182DMedia,Mirabello2019UnderstandingMicroscopy}.
EMT nowadays faces challenge to facilitate designing, synthesizing, and mixing of such media.

\begin{figure}
\centering
\includegraphics{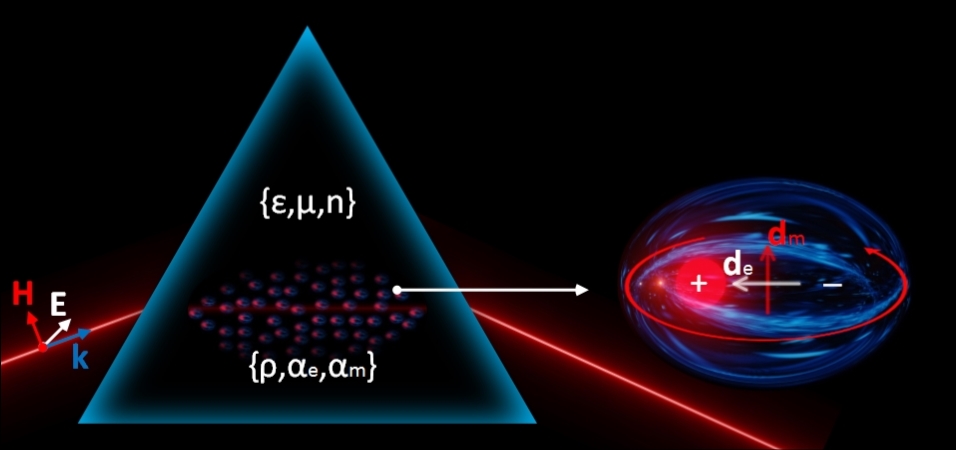}
\caption{\label{fig:M_D}
A medium composed of particles characterized by electric, $\alpha_{\rm e}$, and magnetic, $\alpha_{\rm m}$, polarizabilities. Under external electromagnetic field, electric, $\v d_{\rm e}$, and magnetic, $\v d_{\rm m}$, dipole moments are induced on particles.}
\end{figure}

In this paper we generalize CMR from purely electric materials to materials with both electric and magnetic responses.
The materials considered in this paper is distinguished from the chiral materials~\cite{Sihvola1990ChiralFormula,Lakhtakia1990DiluteSpheres,Tretyakov1995MaxwellInclusions,Weiglhofer1993Maxwell-garnettObjects,Sihvola1992AnalysisMixtures} where the electric polarizability also depends on the magnetic field and the magnetic polarizability also depends on the electric field.
The materials considered in this paper are not required to be chiral.
As could be seen in Fig.~\ref{fig:M_D}, here the electric polarizability only has electric field dependence while the magnetic polarizability only has magnetic field dependence.
The magnetoelectric coupling in electromagnetic media has also been considered in the derivation of constitutive parameters of metamaterials with well defined micro structure and periodicity~\cite{smith_analytic_2010,sozio_generalized_2015,alu_first-principles_2011,smith_composite_2000,Bowen2017}, 
our theory considers randomly distributed sub-wavelength particles.
Similar to previous studies based on the quasicrystalline approximation\cite{Wang2018AchievingParticles,Ding1989EffectivePermittivities}, we have developed the effective medium theory in this work, incorporating the magnetic responses of the constituent particles. Notably, our analysis also encompasses the coupling effect between the electric dipole moment and the magnetic field, as well as the interaction between the magnetic dipole moment and the electric field.

The paper is organized as follows. 
In Sec.~\ref{Theory}, we first provide the CMR for purely electric materials and generalized it to include magnetic response.
In Sec.~\ref{Discussion}, we discuss the interesting physics and materials with negative or near-zero refractive index which merged from the theory.
Finally, we compare the refractive index of the materials calculated by the theory in this paper with the simulated results and the results from CMR.
The SI system of units is used throughout the paper.

\section{Theory}
\label{Theory}
\subsection{The Clausius-Mossotti relation}

Before considering a generalized CMRs, it is expedient to provide a derivation of its well-established representation.
Consider a composite medium consisting of particles with sizes much smaller than the wavelength of the incident illumination. 
Particles are embedded in vacuum and characterized by complex-valued electric dipole polarizability, $\alpha_{\rm e}$.
The polarizability may takes into account finite size effects~\cite{Shalaev1996ElectromagneticComposites}, self-interaction~\cite{deVries1998PointWaves}, and dipole fluctuations~\cite{Barrera1988}.
The electric dipole moment induced on the particles are taken to be independently linear in the total electric field on them: $\v d_{\rm e} = \alpha_{\rm e} \d E$ with higher order multipoles neglected. 
The goal is to replace the particles at the microscopic level with a continuous medium at the macroscopic level, requiring the electric polarization of the medium to be the same after the replacement~\cite{Markel2016IntroductionTutorial,Markel2016MaxwellTutorial}.
The macroscopic expression of the electric polarization is given by

\begin{equation}
\label{eq:P_macro}
\v P = (\varepsilon-1)\varepsilon_{\rm 0}\v E.
\end{equation}
Here $\varepsilon_{\rm 0}$ is the vacuum permittivity, and $\d E$ is the Maxwell's macroscopic electric field in a medium which is averaged over a volume $V_{\rm avg}$ satisfying $1/\rho \ll V_{\rm avg} \ll \lambda^3$ \cite{Markel2016IntroductionTutorial}, where $\rho$ is the number density of the particles/molecules and $\lambda$ is the wavelength of the monochromatic field.
The macroscopic electric field $\v E$ is generated by the incident electric field $\v E_{\rm inc}$ as well as the polarization of the medium~\cite{chew1995waves}

\begin{equation}
\label{eq:VIE_macro}
\v E(\v r) = \v E_{\rm inc}(\v r) +  
\int \omega^2\mu_{\rm 0}\dy G(\v r, \v r')\cdot \v P(\v r') {\rm d}^3  r'.
\end{equation}
Here $\omega$ is the angular frequency of the field and $\mu_{\rm 0}$ is the vacuum permeability. The dyadic Green's function $\dy G$ satisfies the equation
\begin{equation}
\label{eq:Green}
    [\nabla\times\nabla\times - \, \omega^2 \varepsilon_{\rm 0}\mu_{\rm 0}]\dy G(\v r, \v r')=\delta(\v r-\v r') \bar{\bar I}_{3}.
\end{equation}
Here $\nabla\times$ denotes the curl operator acting on $\v r$ and $\bar{\bar I}_{3}$ is an identity tensor. 
The first principle derivation of Eq.~\eqref{eq:VIE_macro} from the Maxwell's equations is given in the supplementary material.
The field $\v E$ in the medium is solved by combining Eq.~\eqref{eq:P_macro} and Eq.~\eqref{eq:VIE_macro}.

The microscopic expression of the electric polarization is given by

\begin{equation}
\label{eq:P_micro}
\v P = \rho \alpha_{\rm e} \d E,
\end{equation}
where $\rho$ is the number density of the constituents.
The microscopic field $\d E$ is calculated by ensemble-averaging, which is justified as follows.
The Poynting vector, describing the energy flow of the electromagnetic field, can be calculated by its time average.
Assuming ergodicity, we replace the time average by an ensemble average~\cite{mishchenko_multiple_2006}.
The ensemble-averaged Poynting vector is decomposed into a coherent flux $\d S_{\rm coh} (\v r) = {\rm Re}[\d E (\v r) \times \d H^* (\v r)]/2$ and an incoherent part given by Eq.~(14) in Ref.~\cite{Mackowski2013DirectParticles}.
Here the coherent electric field $\d E (\v r)$ is calculated by averaging over all realizations of particles \cite{ishimaru_theory_1977} and $\d H$ is the magnetic analogue of $\d E$.

The microscopic field $\d E$ is different from the macroscopic field $\v E$.
When calculating $\d E(\v r)$ at a location $\v r$, we assign a Lorentzian sphere~\cite{lorenz1880ueber,lorentz1881ueber} centered at $\v r$ with a volume limit $V_L < 1/\rho \ll \lambda^3$, where $\lambda$ is the wavelength of the field in the effective medium.
The field at $\v r$ scattered from a particle located within the sphere is taken into account when calculating the polarizability of the particle through self-interaction~\cite{deVries1998PointWaves} and dipole fluctuations~\cite{Barrera1988}, thus, is considered as no contribution to $\d E(\v r)$~\cite{born2013principles}.
The microscopic field $\d E(\v r)$ is thus generated by the incident field and the particles excluded from $V_L$ ~\cite{debye1912theorie,onsager1936electric,Lewin1947,garnett_xii_1904,garnett_vii_1906,bottcher1942theorie,kirkwood1936theory}

\begin{equation}
\label{eq:VIE_micro}
\d E(\v r) = \v E_{\rm inc}(\v r) +  
\oint \omega^2\mu_{\rm 0}\dy G(\v r, \v r')\cdot \v P(\v r') {\rm d}^3  r'.
\end{equation}
Here $\oint...$ denotes an integration over the space of the medium excluding the Lorentzian sphere $V_L$. We subtract Eq.~\eqref{eq:VIE_micro} from Eq.~\eqref{eq:VIE_macro} to have

\begin{equation}
\label{eq:E_diff}
\v E(\v r)-\d E(\v r)
= \int_{V_L} \omega^2\mu_{\rm 0}\dy G(\v r, \v r')\cdot \v P(\v r') {\rm d}^3  r'.
\end{equation}
The dyadic Green's function in free-space $\dy G(\v r, \v r')$ is decomposed into a singular part $-\frac{1}{3\omega^2\varepsilon_{\rm 0}\mu_{\rm 0}}\delta(\v r - \v r')$ and a regular part~\cite{Panasyuk2009} which can be neglected when integrated over the Lorentzian sphere~\cite{Markel2016IntroductionTutorial,chew1995waves,born2013principles,Yaghjian1980}.
The integration with the singular part of the Green's function gives

\begin{equation}
\label{eq:E_diff_2}
\v E(\v r)-\d E(\v r)
= -\frac{\v P(\v r)}{3\varepsilon_{\rm 0}} = -\frac{\varepsilon -1}{3}\v E(\v r).
\end{equation}
Notice that Eq.~\eqref{eq:VIE_macro} is used in the derivation of the second equivalence. Rearranging Eq.~\eqref{eq:E_diff_2} gives

\begin{equation}
\label{eq:E_LFC}
\d E
 = \frac{\varepsilon+2}{3}
\v E,
\end{equation}
which is usually referred to as the local field correction, describing the relation between the microscopic and macroscopic electric fields. Combining Eq.~\eqref{eq:P_micro}, Eq.~\eqref{eq:P_macro}, and Eq.~\eqref{eq:E_LFC}, we have the CMR

\begin{equation}
\label{eq:CMR_E}
\frac{\rho\alpha_{\rm e}}{3\varepsilon_{\rm 0}} = \frac{\varepsilon -1}{\varepsilon + 2},
\end{equation}
which bridges the microscopic electric response of the constituents of the medium on the left hand side and the macroscopic electric response of the medium on the right hand side.

Following the same derivation as presented above, but for a purely magnetic medium, we have the magnetic analogue of the CMR by Poisson~\cite{poisson1825second}:

\begin{equation}
\label{eq:CMR_M}
\frac{\rho\alpha_{\rm m}}{3\mu_{\rm 0}} = \frac{\mu -1}{\mu + 2}.
\end{equation}
CMR, Eq.~\eqref{eq:CMR_E}, is derived and validated in a purely electric medium with $\alpha_{\rm m}=0$ and $\mu=1$. 
Similarly, Poisson's theory, Eq.~\eqref{eq:CMR_M}, is derived in a purely magnetic medium with $\alpha_{\rm e}=0$ and $\varepsilon=1$.
One can approximate $\varepsilon$ by Eq.~\eqref{eq:CMR_E} and $\mu$ by Eq.~\eqref{eq:CMR_M} in dilute media, assuming mutually independent electric and magnetic response~\cite{Sihvola1990ChiralFormula,Lakhtakia1990DiluteSpheres,Tretyakov1995MaxwellInclusions,Weiglhofer1993Maxwell-garnettObjects,Sihvola1992AnalysisMixtures}.
However, in what follows, we claim that in a medium with both strong electric and magnetic responses one should not use Eq.~\eqref{eq:CMR_E} to calculate $\varepsilon$ and use Eq.~\eqref{eq:CMR_M} to calculate $\mu$.
In the derivation of CMR, the microscopic electric field in Eq.~\eqref{eq:VIE_macro} and the macroscopic electric field in Eq.~\eqref{eq:VIE_micro} are contributed by the field scattered from the electric polarization of the medium.
However, when induced by the magnetic field in the medium, the magnetic polarization also contributes to the electric field distribution in the medium.
The mutual dependence of the electric and magnetic field missing in the conventional effective medium theories leads us to the revisit and generalization of CMR in the next section.

\subsection{The generalized Clausius-Mossotti relation for media with both electric and magnetic response}

Consider the EMT of a composite medium with particles which also response to magnetic field characterized by magnetic dipole polarizability $\alpha_{\rm m}$.
The equivalence of the microscopic and macroscopic polarization gives

\begin{equation}
\label{eq:P_em}
\begin{bmatrix}
(\varepsilon-1)\varepsilon_{\rm 0}\v E \\
(\mu-1)\mu_{\rm 0}\v H
\end{bmatrix}
 =
\begin{bmatrix}
\v P \\
\v M
\end{bmatrix}
= \rho
\begin{bmatrix}
\alpha_{\rm e} \d E \\
\alpha_{\rm m} \d H
\end{bmatrix}
,
\end{equation}
where $M$ is the magnetic polarization.

Similar to Eq.~\eqref{eq:VIE_macro}, the macroscopic electromagnetic field is generated by the incident field and the polarization of the medium

\begin{equation}
\label{eq:VIE_em_macro}
\begin{bmatrix}
\v E(\v r) \\
\v H(\v r) 
\end{bmatrix}
 =
\begin{bmatrix}
\v E_{\rm inc}(\v r) \\
\v H_{\rm inc}(\v r) 
\end{bmatrix}
+  
\int
\overline{\v G}(\v r, \v r')
\cdot
\begin{bmatrix}
\v P(\v r') \\
\v M(\v r')
\end{bmatrix}
{\rm d}^3  r'.
\end{equation}
Here the macroscopic field $\v H$ is the magnetic analogue of $\v E$, $\v H_{\rm inc}$ is the incident magnetic field, and
\begin{equation}
\label{eq:G}
\overline{\v G}(\v r, \v r')
=
\begin{bmatrix}
\omega^2\mu_{\rm 0}\dy G(\v r, \v r') &  i\omega\dy G(\v r , \v r') \cdot \nabla' \times\\
-i\omega\dy G(\v r , \v r') \cdot \nabla' \times & \omega^2\varepsilon_{\rm 0}\dy G(\v r, \v r')
\end{bmatrix}
.
\end{equation}
A detailed derivation from first principle (the Maxwell's equations) of Eq.~\eqref{eq:VIE_em_macro} is given by Ref.~\cite{chew1995waves,sun2009novel} which is summarized in the supplemental material.
The terms in Eq.~\eqref{eq:VIE_em_macro} with the diagonal elements in $\overline{\v G}(\v r, \v r')$ are the electric field generated by the electric polarization (in the first line) and the magnetic field generated by the magnetic polarization (in the second line).
In a purely electric material where $\v M = 0$, the first line of Eq.~\eqref{eq:VIE_em_macro} becomes Eq.~\eqref{eq:VIE_macro}, which means that there is only electric-electric interaction between particles.
In a material with both electric and magnetic responses, the electric and magnetic fields are mutually dependent.
The mutual dependence is given by the terms in Eq.~\eqref{eq:VIE_em_macro} with the non-diagonal elements in $\overline{\v G}(\v r, \v r')$ which are the electric field generated by the magnetic polarization (in the first line) and the magnetic field generated by the electric polarization (in the second line)~\cite{Yaghjian1980}.
The inclusion of the mutual dependence of the electric and magnetic fields in CMR leads to the novel theory in this paper.

Similar as in the derivation of the original CMR, with the introduction of the Lorentzian sphere, we have

\begin{equation}
\label{eq:VIE_em_micro}
\begin{bmatrix}
\d E(\v r) \\
\d H(\v r) 
\end{bmatrix}
 =
\begin{bmatrix}
\v E_{\rm inc}(\v r) \\
\v H_{\rm inc}(\v r) 
\end{bmatrix}
+  
\oint
\overline{\v G}(\v r, \v r')
\cdot
\begin{bmatrix}
\v P(\v r') \\
\v M(\v r')
\end{bmatrix}
{\rm d}^3  r'.
\end{equation}
A detailed derivation of Eq.~\eqref{eq:VIE_em_micro} is given in Ref.~\cite{Wang2021ClusteringParticles}. The mutual dependence terms are also justified in Ref.~\cite{Yaghjian1980}
\footnote{The off-diagonal terms in Eq.~\eqref{eq:VIE_em_micro} are consistent with \cite[Eq.(5a)]{Yaghjian1980}, but with the scalar Green’s function replaced by the dyadic Green’s function, which is justified by taking a curl of \cite[Eq.(2b)]{Yaghjian1980} to get $\nabla \times \nabla \times \textbf{B} - k^2 \textbf{B} = \mu_{\rm 0} \nabla\times \textbf{J}$.
Notice that conventional \cite[Eq.(3a)]{Yaghjian1980} is derived upon using $\nabla\times\nabla\times \textbf{B}=\nabla(\nabla \cdot \textbf{B})-\nabla^2 \textbf{B}$ and $\nabla \cdot \textbf{B}=0$. 
The latter is no longer true in case of the medium containing magnetic sources. 
Keeping $\nabla\times\nabla\times \textbf{B}$ instead of $-\nabla^2 \textbf{B}$ gives the off-diagonal terms in Eq.~\eqref{eq:VIE_em_micro}.}.


Following the conventional derivation of the CMR, we subtract Eq.~\eqref{eq:VIE_em_micro} from Eq.~\eqref{eq:VIE_em_macro}:

\begin{equation}
 \begin{aligned}
\label{eq:EH_diff}
\begin{bmatrix}
\v E(\v r) \\
\v H(\v r) 
\end{bmatrix}
 -
\begin{bmatrix}
\d E(\v r) \\
\d H(\v r) 
\end{bmatrix}
= \int_{V_L}
\overline{\v G}(\v r, \v r')\cdot
\begin{bmatrix}
\v P(\v r') \\
\v M(\v r')
\end{bmatrix}
{\rm d}^3  r'
= -\frac{\varepsilon\mu -1}{3}
\begin{bmatrix}
\v E(\v r) \\
\v H(\v r)
\end{bmatrix}
,
\end{aligned}
\end{equation}

\noindent
where we have used Eq.~\eqref{eq:P_em}, Eq.~\eqref{eq:G}, and source-free Maxwell equations: $\nabla \times \v E = i\omega \mu \mu_{\rm 0} \v H$, $\nabla \times \v H = -i\omega \varepsilon \varepsilon_{\rm 0} \v E$. 
Again, the integration including the regular part of the dyadic Green's function over $V_L$ is neglected.
The relation between the microscopic and macroscopic fields is obtained by rearranging Eq.~\eqref{eq:EH_diff}:

\begin{equation}
\label{eq:EH_LFC}
\begin{bmatrix}
\d E \\
\d H 
\end{bmatrix}
 = \frac{\varepsilon\mu+2}{3}
\begin{bmatrix}
\v E \\
\v H 
\end{bmatrix}.
\end{equation}
This generalized local field correction~\cite{Dolgaleva2012Local-fieldMaterials}, together with Eq.~\eqref{eq:P_em}, result in new self-consistent equations
    
\begin{subnumcases}{\label{eq:CM_rev}}
    \; \label{eq:CM_rev_1}
\frac{\rho\alpha_{\rm e}}{3\varepsilon_{\rm 0}} = \frac{\varepsilon -1}{\varepsilon\mu + 2} \\
    \; \label{eq:CM_rev_2}
\frac{\rho\alpha_{\rm m}}{3\mu_{\rm 0}} = \frac{\mu -1}{\varepsilon\mu + 2}.
\end{subnumcases}
Eqs.~(\ref{eq:CM_rev}) generalize the CMR, which is valid only in purely electric materials, to materials with both electric and magnetic responses, bridging the microscopic properties of the particles (on the left hand sides) with the macroscopic properties of the effective medium (on the right hand sides).
In a purely electric material with $\alpha_{\rm m}=0$, Eq.~(\ref{eq:CM_rev_2}) gives $\mu = 1$.
Thus Eq.~(\ref{eq:CM_rev_1}) recovers the CMR, Eq.~(\ref{eq:CMR_E}).
Similarly, Poisson's theory is recovered by Eqs.~(\ref{eq:CM_rev_2}) in purely magnetic materials.
Thus, our Eqs.~(\ref{eq:CM_rev}) are consistent with conventional EMTs, and, most importantly, fill the gap between EMTs for purely electric and for purely magnetic materials.

The CMR, Eq.~(\ref{eq:CMR_E}), along side with its subsequent generalizations~\cite{garnett_xii_1904,garnett_vii_1906,onsager1936electric,debye1912theorie,poisson1825second,bottcher1942theorie,kirkwood1936theory,Lewin1947,Sharma2000StudiesParticles,McPhedran1997DynamicFormula}, bridges the electric microscopic response of the particles with the electric macroscopic response of the medium and Possion's theory, Eq.~(\ref{eq:CMR_M}), bridges the magnetic-magnetic responses.
Our theory, Eqs.~(\ref{eq:CM_rev}), further includes the mutual dependence in CMR.
We can observe from Eqs.~(\ref{eq:CM_rev}) that now the permittivity of the medium depends also on the magnetic polarizability of the particles $\alpha_{\rm m}$ and the permeability depends also on the electric polarizability $\alpha_{\rm e}$.
This mutual dependence is a result of including the off-diagonal terms of $\overline{\v G}(\v r, \v r')$ in Eq.~(\ref{eq:G}), which correspond to the electric field generated by the magnetic polarization (the $2$nd term in the first line) and the magnetic field generated by the electric polarization (the $1$st term in the second line).
The mutual dependence brought up in this paper is resulted from the macroscopic Maxwell's equation (the derivation could be found in the supplementary material) thus put no constrains on the microscopic structure of the materials, not requiring the materials to be chiral.

\section{Discussion}
\label{Discussion}

In this section, we first reform Eq.~\eqref{eq:CM_rev} and provide the solution of the macroscopic properties of materials in terms of the microscopic properties.
Then, we will analyze the new physics merged from the solutions, that in one material there are two sets of solutions of refractive indices.
We will also find surprising properties of materials merged from this theory, which are near-zero-index materials and double-negative materials.

\subsection{Solution of the macroscopic properties and the two branches}

Solving Eqs.~(\ref{eq:CM_rev}), we find $\varepsilon$ and $\mu$ in terms of the microscopic properties:

\begin{align}
\label{eq:epsilon_ind}
\varepsilon = \frac{ 1-\beta_{\rm e} + \beta_{\rm m} + \gamma }{2\beta_{\rm m}},
\\
\label{eq:mu_ind}
\mu = \frac{ 1+\beta_{\rm e} - \beta_{\rm m} + \gamma }{2\beta_{\rm e}},
\end{align}
where
\begin{equation}
\label{eq:gamma}
\gamma \equiv \pm \sqrt{\beta_{\rm e}^2+\beta_{\rm m}^2-10\beta_{\rm e}\beta_{\rm m}-2\beta_{\rm e}-2\beta_{\rm m}+1},
\end{equation}

\noindent
and $\beta_{\rm e} \equiv \rho\alpha_{\rm e}/(3\varepsilon_{\rm 0})$, $\beta_{\rm m} \equiv \rho\alpha_{\rm m}/(3\mu_{\rm 0})$. 
The branch choice in Eq.~\eqref{eq:gamma} raises the tantalizing possibility that the macroscopic material may support two sets of macroscopic material parameters $\varepsilon$, $\mu$, $n$ simultaneously.
This emerges not as an effect of anisotropy, but rather the propagation of two distinct, generally inhomogeneous waves with different complex-valued wavenumbers at the same frequency.

While we predict that both sets of macroscopic material parameters are always present, in cases commonly encountered media, only one is practically observable.
We should notice that most materials are non- or weakly-magnetic, i.e. $\beta_{\rm m} \ll {\rm min}\{\beta_{\rm e}, 1\}$ or $\beta_{\rm m} \to 0$. 
The two solutions of the squares of refractive indices are given by

\begin{equation}
\label{eq:n_appr}
n^2 = \frac{(1-\beta_{\rm e} \pm |1-\beta_{\rm e}|)+f(\beta_{\rm e})\beta_{\rm m}+O(\beta_{\rm m}^2)}{2\beta_{\rm e}\beta_{\rm m}}.
\end{equation}

\noindent
With $\beta_{\rm m} \to 0$, the solutions of the refractive index in one of the two branches will be infinite, leading to a fast-oscillating and high-loss propagation of light in the medium, summarized as

\begin{numcases}{}
\label{eq:n_+_inf}
n_{\rm +} \to \infty, \; \rm as \; \beta_{\rm m} \to 0, \, \beta_{\rm e}\ge 1
\\[14pt]
\label{eq:n_-_inf}
n_{\rm -} \to \infty, \; \rm as \; \beta_{\rm m} \to 0, \, \beta_{\rm e}<1
.
\end{numcases}

\noindent
This results indicate that only one of the two branches could be observed or measured in a non- or weak-magnetic medium. 
The other branch of solutions leads to an infinite refractive index. 
The electromagnetic wave associated with this refractive index is thus not observable. 

However, in media consisting of particles/elements with simultaneously strong electric and magnetic polarizabilities, both electromagnetic waves corresponding to the two branches of solutions should be observed.
One of such types of media is metamaterials, which are made of periodically aligned artificial subwavelength structures that can be polarized by both electric and magnetic fields.
It is well-known that when an interface between the metamaterial and the free-space is illuminated by an incident electromagnetic plane wave, a negatively refracted wave could be found in the metamaterial.
While the observation of the negatively refracted wave is widely reported in experiments, it is less mentioned that usually a positively refracted wave coexists in the metamaterial~\cite{Couture2010ExperimentalMetamaterial,Ozbay2008NegativeMetamaterial,Gao2013MeasurementSample,Chen2005NegativeMetamaterial,RAN2003Left-handedVerifications,Aydin2005ObservationMetamaterials,Huangfu2004ExperimentalPatterns,Wang2010ExperimentalTerahertz,Chen2006MetamaterialWaves,Beruete2009NegativeSlab,Xi2009ExperimentalMetamaterial,Ozbay2007MetamaterialsSimulations,Smith2004MetamaterialsIndex,Dong2005NumericalMetamaterials,Shelby2001ExperimentalRefraction,Sun2010LowArrays}.
It is also worth mentioning that in the metamaterial consisting of structures with weak-magnetic response, the positively refracted wave is not observed~\cite{Imhof2009ExperimentalMetamaterial}.
The positively refracted waves in the previous references are usually tuned to be less dominant than the negatively refracted waves because the negative refraction was the focus of those works.
However, both waves could be dominant in a medium, depending on the frequency of the field or the size parameters of the periodic structures~\cite{Smith2004EnhancedMetamaterial}.

\begin{figure}
\centering
\includegraphics{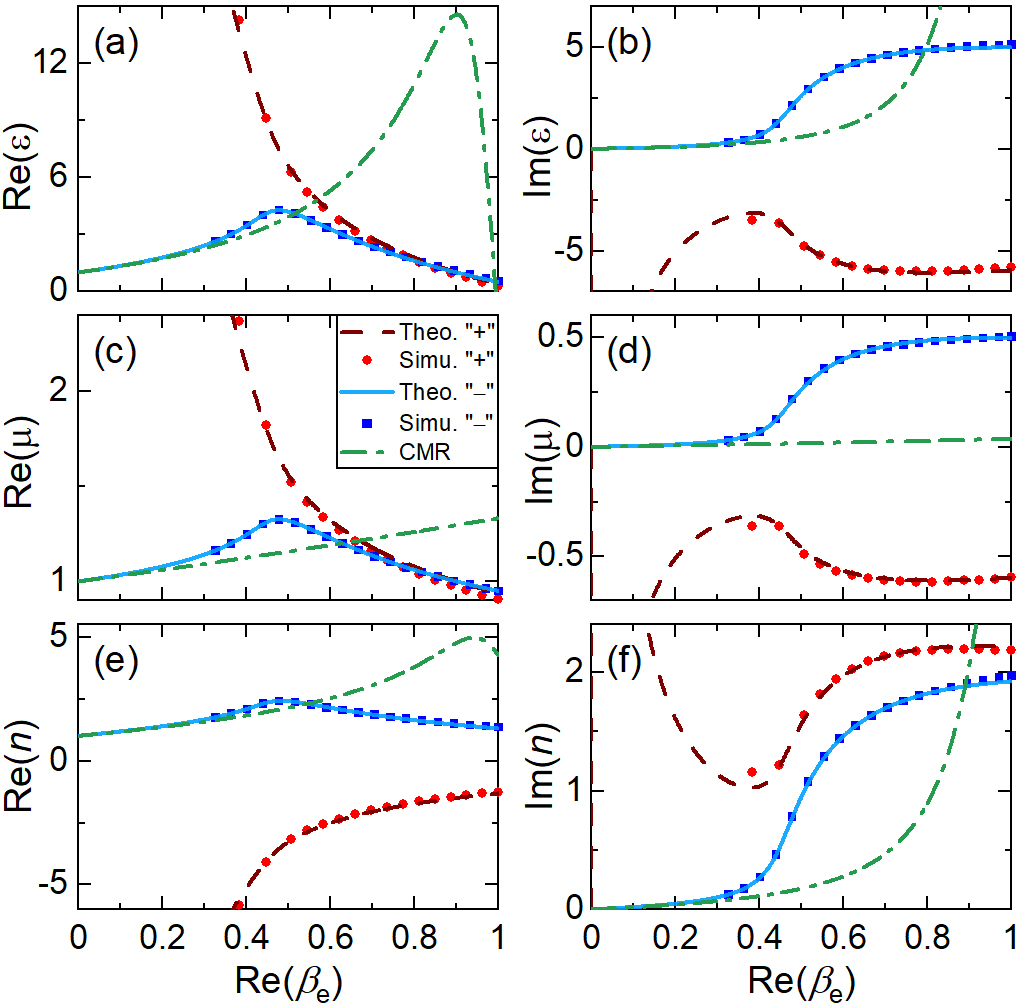}
\caption{\label{fig:6curve}
The theoretical and simulated macroscopic $\varepsilon$, $\mu$, and $n$ varying with microscopic $\beta_{\rm e}$ and $\beta_{\rm m}$. The ratios are assigned as ${\rm Im}(\beta_{\rm e})/{\rm Re}(\beta_{\rm e})={\rm Im}(\beta_{\rm m})/{\rm Re}(\beta_{\rm m})=\beta_{\rm m}/\beta_{\rm e}=0.1$. The conventional CMR results are plotted in green dotted-dash line as a comparison.}
\end{figure}

\subsection{Numerical verification and comparison with the conventional CMR}

The generalized CMR is verified by simulations with our recently developed numerical technique, the clustering diffused-particle method~\cite{Wang2021ClusteringParticles}. 
We generate $4\times 10^{10}$ particles randomly distributed in a cube with a side length of $4.2$ wavelengths.
We simulate the scattering of a plane wave from a large collection of particles and then extract the effective planewave-like behavior in order to find the effective macroscopic parameters of the medium. 
An iterative solution of a generalized Foldy-Lax equation provides the field distribution and effective macroscopic responses of the medium: $\varepsilon$, $\mu$, and $n$~\cite[Sec.III.A]{Wang2021ClusteringParticles}.
Both solutions of Eq.~\eqref{eq:epsilon_ind} and Eq.~\eqref{eq:mu_ind} agree well with numerical simulations, as shown in Fig.~\ref{fig:6curve}. 
As it might be expected, for the weak electric response, i.e., for small $\beta_e$, one of the branches coincides with the conventional CMR.
This is clearly observed in Fig.~\ref{fig:6curve} for the ``--'' solution for $\beta_e\lesssim 0.3$.
The ``+'' branch, however, yields divergent values for $|\varepsilon|$, $|\mu|$ and $|n|$ as $\beta_e\to 0$.
Thus waves associated with this wavenumber oscillate and decay on very short length scales.
This may explain why this second electromagnetic state of the medium has not been reported in weak- or non- magnetic materials.
Both solutions should be observable when both $\beta_{\rm e}$ and $\beta_{\rm m}$ are appreciable.
For example, $\beta_e\gtrsim 0.3$, and $\beta_{\rm m} = 0.1\beta_{\rm e}$, we find not only a discrepancy between the conventional CMR and the ``--'' branch emerges, but the ``+'' branch yields finite values of $\{ \varepsilon, \mu, n\}$.
It is worth noting, the ``--'' and ``+'' branches provide the same values for ${\rm Re}(\varepsilon)$ and ${\rm Re}(\mu)$ at ${\rm Re}(\beta_e)\to 1$.

\subsection{Engineering media with exotic response}

\begin{figure}[t!]
\centering
\includegraphics{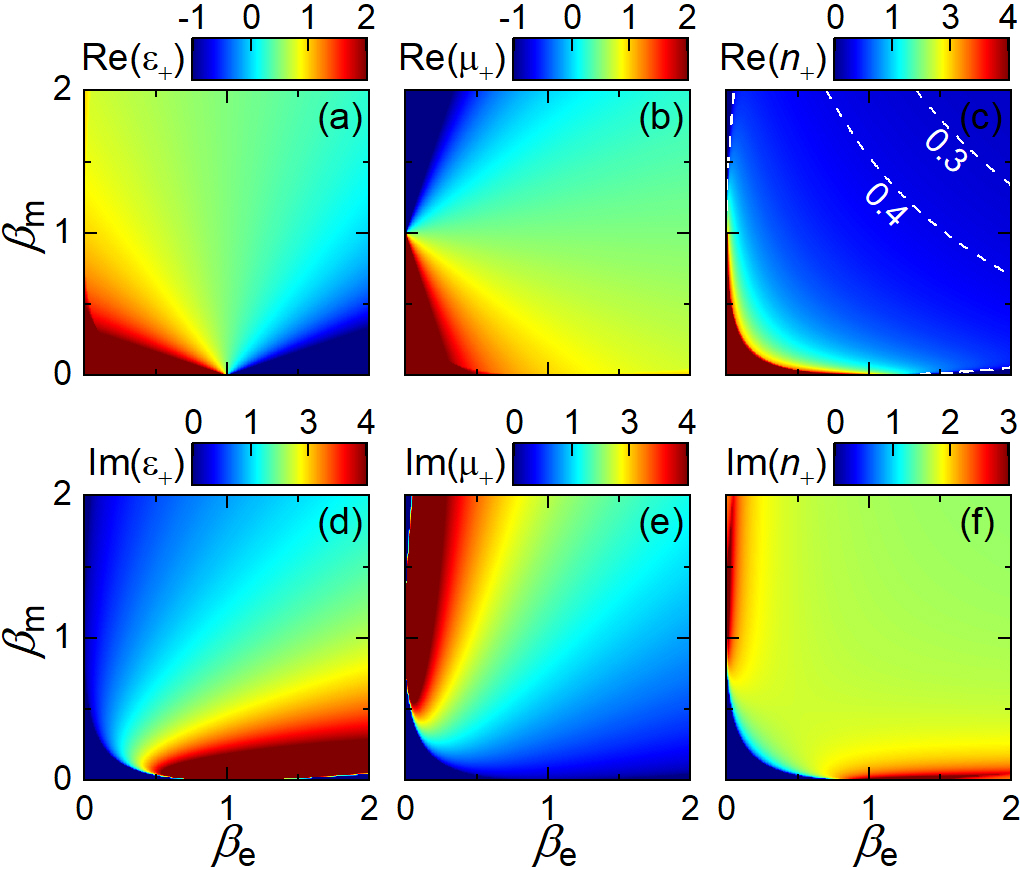}
\caption{\label{fig:emn+}
Permittivity, permeability and refractive index for ``+'' branch of $\gamma$, Eq.~\eqref{eq:gamma}: (a) ${\rm Re}( \varepsilon_{\rm +} )$, (b) ${\rm Re}( \mu_{\rm +} )$, (c) ${\rm Re}( n_{\rm +} )$, (d) ${\rm Im}( \varepsilon_{\rm +} )$, (e) ${\rm Im}( \mu_{\rm +} )$, (f) ${\rm Im}( n_{\rm +} )$.
For a specific material, $\beta_{\rm e}$ and $\beta_{\rm m}$ can be varied by changing the concentration $\rho$. 
Under standard conditions (i.e., room temperature and 1 atm pressure) most chemical elements and compounds~\cite{schwerdtfeger_table_2015} are characterized by $0<|\beta_{\rm e}|<3$. 
Magnetic counterpart $\beta_{\rm m}$ is typically much smaller than $1$~\cite{Pulido-Mancera2017PolarizabilityModelingb,Rockstuhl2007TheApproach,Jylha2006ModelingSpheres}.}
\end{figure}

\begin{figure}[ht]
\centering
\includegraphics{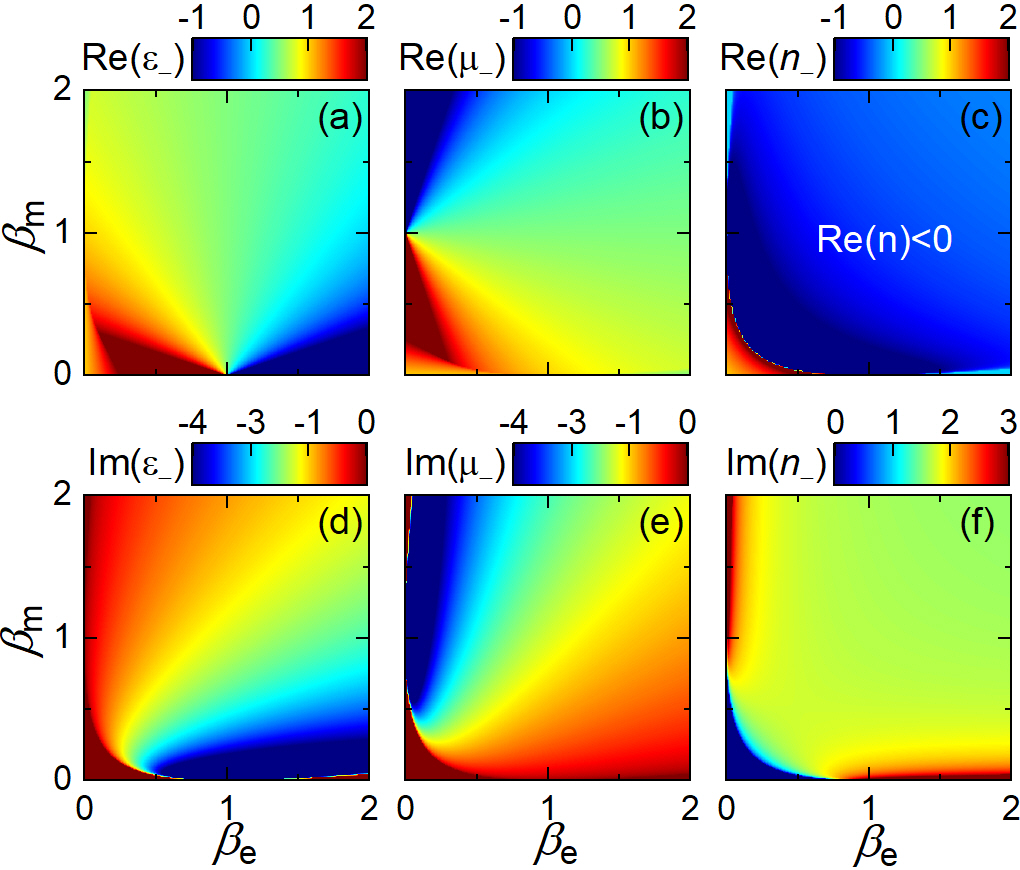}
\caption{\label{fig:emn-}
Same as in Fig.~\ref{fig:emn+}, but for the ``--'' branch, Eq.~\eqref{eq:gamma}.
Arrows along the $x$-axis (CM) in (a), (c), (d), (f) correspond to the range of applicability of conventional Clausius-Mossotti theory~\cite{clausius1879mechanical,mossotti1850discussione}, while arrows along the $y$-axis (P) in (b), (c), (e), (f) denote the respective range for Poisson's theory~\cite{poisson1825second}.
}
\end{figure}

Eq.~\eqref{eq:CM_rev} allows us to engineer the macroscopic electromagnetic responses of the materials by manipulating the microscopic responses of the particles.
Because of the mutual dependence of the electric and magnetic fields, a variation of either $\alpha_{\rm e}$ or $\alpha_{\rm m}$ would change the values of $\varepsilon$ and $\mu$, which provides us more flexibility when tailoring optimized macroscopic electromagnetic responses.
The dependence of $\varepsilon$, $\mu$ and $n$ on both $\beta_{\rm e}$ and $\beta_{\rm m}$ are shown in \rfig{emn+} and \rfig{emn-}, for ``+'' and ``--'' solution respectively.
Notice that $x$-axis (i.e., $\alpha_{\rm m}=0$) and $y$-axis (i.e., $\alpha_{\rm e}=0$) on these plots represent Clausius-Mossotti's and Poission's theories, respectively.
The refractive indices $n$ are calculated as $n=\pm\sqrt{\varepsilon\mu}$.
The strategy of choosing the sign of $n$ is provided in \cite{mccall_negative_2002}.
Remarkably, the argument under the square root in Eq.~\eqref{eq:gamma} can be negative even when all the $\beta_{\rm e}$ and $\beta_{\rm m}$ are real and positive and so macroscopic characteristics $\varepsilon$, $\mu$ and $n$ can be complex-valued. 
Even though the particles are lossless, light propagating through a collection of particles is scattered out of the direction of the propagation, which results in an apparent loss of energy~\cite{Bohren1998}.


\begin{figure}
\centering
\includegraphics[width=3in]{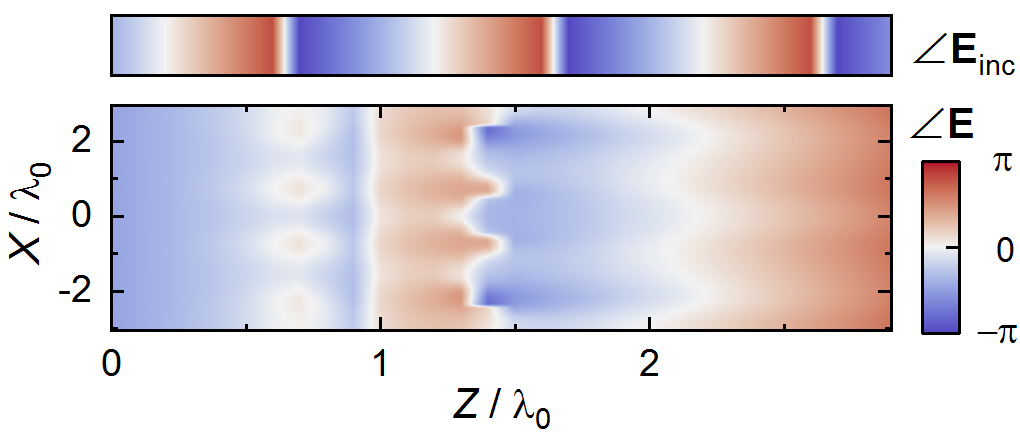}
\caption{\label{fig:Near0}
The phase of the electric component of the electromagnetic field propagating in a near-zero-index medium with microscopic properties $\beta_{\rm e}=0.0323+1.8497i$ and $\beta_{\rm m}=0.3+0.96i$. 
According to Eq.~\eqref{eq:epsilon_ind} and Eq.~\eqref{eq:mu_ind}, the effective refractive index is $n=0.179 + 1.889i$.
The field is simulated~\cite{Wang2021ClusteringParticles} in a cube with side length $12\lambda_0$ and sampled in a depth of $3\lambda_0$.}
\end{figure}

Notice that although the derivation of the generalized CMR follows the derivation of the conventional CMR dated back to the 19th century, it predicts the existences of the materials with uncommon electromagnetic properties which are discovered in the 21st century, which is, in general, not possible with conventional CMR.
The double-negative materials could be found in \rfig{emn-}(c), corresponding to materials consist of particles/elements with both electric and magnetic responses.
Near-zero-index materials~\cite{Liberal2017TheTechnologies} can also be designed using Eq.~\eqref{eq:epsilon_ind}.
\rfig{Near0} shows the phase of the electric component of the electromagnetic field propagating through such a medium engineered by optimizing the values of $\beta_{\rm e}$ and $\beta_{\rm m}$ to yield ${\rm Re}(n)=0.179$ when $\beta_{\rm e}=0.0323+1.8497i$ and $\beta_{\rm m}=0.3+0.96i$.
Exotic near-zero-index materials might be generated by doping with atoms, molecules, or metamaterials~\cite{Liberal2017PhotonicMedia}.
These results point a way towards engineering new indices of refraction by exploring a parameter space of $\beta_{\rm e}$ and $\beta_{\rm m}$ through maps as shown in \rfig{emn+} and \rfig{emn-} and seeking constituent materials with appropriate combinations of polarizabilities.

\section{Conclusion}

We have shown that if the constituents are both magnetic and electric, within a generalized CMR, the resultant index of refraction can become anomalously large, go to zero or become negative. 
Thus, Eq.~\eqref{eq:epsilon_ind}-Eq.~\eqref{eq:gamma} provide the means for the design of materials with exotic properties.
We suggest that all-dielectric nanocolloids~\cite{Hinamoto2020ColloidalMetafluids} and split-ring resonator~\cite{smith_composite_2000}, exhibiting both strong electric and magnetic responses, may serve as promising building blocks for such materials.
Our results have implications for other generalizations of mixing rules and other homogenizations~\cite{Sihvola1999} including, but not limiting to, non-linear effects~\cite{gehring_observation_2006,bigelow_superluminal_2003,wang_gain-assisted_2000}, permanent dipole moment~\cite{debye1912theorie}, temperature-dependence~\cite{reitz2008foundations}, and constituents with particular shape and volume~\cite{garnett_xii_1904,garnett_vii_1906,Alu2005PolarizabilitiesLayers,Werdehausen2018DesignNanocomposites}.
It should be noted that if the resultant index of refraction reaches an excessively high value, leading to very short wavelengths inside the medium, it can exceed the volume limit of the individual particles, expressed as $V_L < 1/\rho \ll \lambda^3$. Consequently, the effective medium theory becomes inadequate under such circumstances. This theory is applicable only when the wavelength within the effective medium is significantly larger than the dimensions of the composites.
In instances where a significantly large resultant index of refraction occurs, it may potentially impact the accuracy of the results presented, to some extent.

We thank Prof. Vadim A. Markel from the University of Pennsylvania and Prof. John C. Schotland from Yale University for fruitful discussions.

\bibliographystyle{iopart-num}
\bibliography{Mendeley_Lang,zo}

\providecommand{\newblock}{}
\begin{thebibliography}{100}
\expandafter\ifx\csname url\endcsname\relax
  \def\url#1{{\tt #1}}\fi
\expandafter\ifx\csname urlprefix\endcsname\relax\def\urlprefix{URL }\fi
\providecommand{\eprint}[2][]{\url{#2}}

\bibitem{Sihvola1999}
Sihvola A~H 1999 {\em {Electromagnetic mixing formulas and applications}\/}
  (London : Institution of Electrical Engineers)

\bibitem{Choy2016}
Choy T~C 2016 {\em {Effective medium theory: principles and applications}\/}
  2nd ed (Oxford University Press)

\bibitem{Markel2016MaxwellTutorial}
Markel V~A 2016 {\em Journal of the Optical Society of America A\/} {\bf 33}
  2237

\bibitem{Markel2016IntroductionTutorial}
Markel V~A 2016 {\em Journal of the Optical Society of America A\/} {\bf 33}
  1244

\bibitem{Aspnes2011}
Aspnes D~E 2011 {\em Thin Solid Films\/} {\bf 519} 2571--2574

\bibitem{Mishchenko2016ApplicabilityParticles}
Mishchenko M~I, Dlugach J~M and Liu L 2016 {\em Journal of Quantitative
  Spectroscopy and Radiative Transfer\/} {\bf 178} 284--294

\bibitem{mossotti1850discussione}
Mossotti O~F 1850 {\em Mem. Soc. Ital\/} {\bf 14} 49

\bibitem{clausius1879mechanical}
Clausius R 1879 {\em {The mechanical theory of heat}\/} (Macmillan)

\bibitem{lorenz1880ueber}
Lorenz L 1880 {\em Annalen der Physik\/} {\bf 247} 70--103

\bibitem{lorentz1881ueber}
Lorentz H~A 1881 {\em Annalen der physik\/} {\bf 248} 127--136

\bibitem{poisson1825second}
Poisson S~D 1825 {\em {Second m{\'{e}}moire sur la th{\'{e}}orie du
  magn{\'{e}}tisme}\/} (Imprimerie royale)

\bibitem{debye1912theorie}
Debye P 1912 {\em Annalen der Physik\/} {\bf 344} 789--839

\bibitem{onsager1936electric}
Onsager L 1936 {\em Journal of the American Chemical Society\/} {\bf 58}
  1486--1493

\bibitem{Lewin1947}
Lewin L 1947 {\em Journal of the Institution of Electrical Engineers-Part III:
  Radio and Communication Engineering\/} {\bf 94} 65--68

\bibitem{garnett_xii_1904}
Garnett J~M 1904 {\em Philosophical Transactions of the Royal Society of
  London. Series A, Containing Papers of a Mathematical or Physical
  Character\/} {\bf 203} 385--420

\bibitem{garnett_vii_1906}
Garnett J~M 1906 {\em Philosophical Transactions of the Royal Society of
  London. Series A, Containing Papers of a Mathematical or Physical
  Character\/} {\bf 205} 237--288

\bibitem{bottcher1942theorie}
B{\"{o}}ttcher C~J~F 1942 {\em Physica\/} {\bf 9} 937--944

\bibitem{kirkwood1936theory}
Kirkwood J~G 1936 {\em The Journal of Chemical Physics\/} {\bf 4} 592--601

\bibitem{McPhedran1997DynamicFormula}
McPhedran R, Poulton C, Nicorovici N and Movchan A 1997 {\em Physica A:
  Statistical Mechanics and its Applications\/} {\bf 241} 179--182

\bibitem{Sharma2000StudiesParticles}
Sharma R and Sihvola A 2000 {\em Radio Science\/} {\bf 35} 83--96

\bibitem{MohadjerBeromi2019SynthesisReactions}
Mohadjer~Beromi M, Brudvig G~W, Hazari N, Lant H~M~C and Mercado B~Q 2019 {\em
  Angewandte Chemie International Edition\/} {\bf 58} 6094--6098

\bibitem{Martel2014High-resolutionChemistry}
Martel L, Magnani N, Vigier J~F, Boshoven J, Selfslag C, Farnan I, Griveau J~C,
  Somers J and Fangh{\"{a}}nel T 2014 {\em Inorganic Chemistry\/} {\bf 53}
  6928--6933

\bibitem{Luchinat2012Solid-stateRestraints}
Luchinat C, Parigi G, Ravera E and Rinaldelli M 2012 {\em Journal of the
  American Chemical Society\/} {\bf 134} 5006--5009

\bibitem{Knight2012StructureNMR}
Knight M~J, Pell A~J, Bertini I, Felli I~C, Gonnelli L, Pierattelli R, Herrmann
  T, Emsley L and Pintacuda G 2012 {\em Proceedings of the National Academy of
  Sciences\/} {\bf 109} 11095--11100

\bibitem{Estephane2010AState}
Estephane J, Groppo E, Vitillo J~G, Damin A, Gianolio D, Lamberti C, Bordiga S,
  Quadrelli E~A, Basset J~M, Kervern G, Emsley L, Pintacuda G and Zecchina A
  2010 {\em The Journal of Physical Chemistry C\/} {\bf 114} 4451--4458

\bibitem{Flambard2009RevisitingO}
Flambard A, K{\"{o}}hler F~H and Lescou{\"{e}}zec R 2009 {\em Angewandte
  Chemie\/} {\bf 121} 1701--1704

\bibitem{Pell2019ParamagneticState}
Pell A~J, Pintacuda G and Grey C~P 2019 {\em Progress in Nuclear Magnetic
  Resonance Spectroscopy\/} {\bf 111} 1--271

\bibitem{Vaidya2018SubstitutedPhenomenon}
Vaidya S, Shukla P, Tripathi S, Rivi{\`{e}}re E, Mallah T, Rajaraman G and
  Shanmugam M 2018 {\em Inorganic Chemistry\/} {\bf 57} 3371--3386

\bibitem{Adam2018CatalyticReactions}
Adam M~S~S 2018 {\em Applied Organometallic Chemistry\/} {\bf 32} e4234

\bibitem{Gao2018StrongFluids}
Gao R, Zhang Q, Xu Z, Wang Z, Cai W, Chen G, Deng X, Cao X, Luo X and Fu C 2018
  {\em Nanoscale\/} {\bf 10} 11750--11759

\bibitem{Dou2022Two-dimensionalSkyrmions}
Dou K, Du W, Dai Y, Huang B and Ma Y 2022 {\em Physical Review B\/} {\bf 105}
  205427

\bibitem{Vaz2010MagnetoelectricStructures}
Vaz C~A~F, Hoffman J, Ahn C~H and Ramesh R 2010 {\em Advanced Materials\/} {\bf
  22} 2900--2918

\bibitem{Spaldin2019AdvancesMultiferroics}
Spaldin N~A and Ramesh R 2019 {\em Nature Materials\/} {\bf 18} 203--212

\bibitem{Mundy2016AtomicallyMultiferroic}
Mundy J~A, Brooks C~M, Holtz M~E, Moyer J~A, Das H, R{\'{e}}bola A~F, Heron
  J~T, Clarkson J~D, Disseler S~M, Liu Z, Farhan A, Held R, Hovden R, Padgett
  E, Mao Q, Paik H, Misra R, Kourkoutis L~F, Arenholz E, Scholl A, Borchers
  J~A, Ratcliff W~D, Ramesh R, Fennie C~J, Schiffer P, Muller D~A and Schlom
  D~G 2016 {\em Nature\/} {\bf 537} 523--527

\bibitem{Nan1993}
Nan C~W 1993 {\em Progress in Materials Science\/} {\bf 37} 1--116

\bibitem{Nan2008}
Nan C~W, Bichurin M~I, Dong S, Viehland D and Srinivasan G 2008 {\em Journal of
  Applied Physics\/} {\bf 103} 31101

\bibitem{Jiang2007MagnetoelectricSintering}
Jiang Q~H, Shen Z~J, Zhou J~P, Shi Z and Nan C~W 2007 {\em Journal of the
  European Ceramic Society\/} {\bf 27} 279--284

\bibitem{Yang2022ObservationMetamaterial}
Yang W, Liu Q, Wang H, Chen Y, Yang R, Xia S, Luo Y, Deng L, Qin J, Duan H and
  Bi L 2022 {\em Nature Communications\/} {\bf 13} 1719

\bibitem{Louis2018AModel}
Louis D, Lacour D, Hehn M, Lomakin V, Hauet T and Montaigne F 2018 {\em Nature
  Materials\/} {\bf 17} 1076--1080

\bibitem{Dudek2019ImpactMetamaterials}
Dudek K~K, Wolak W, Gatt R and Grima J~N 2019 {\em Scientific Reports\/} {\bf
  9} 3963

\bibitem{Zhao2019IntegratingMetadevices}
Zhao X, Duan G, Li A, Chen C and Zhang X 2019 {\em Microsystems {\&}
  Nanoengineering\/} {\bf 5} 5

\bibitem{Duan2019BoostingMetamaterials}
Duan G, Zhao X, Anderson S~W and Zhang X 2019 {\em Communications Physics\/}
  {\bf 2} 35

\bibitem{Koschny2004}
Koschny T, Kafesaki M, Economou E~N and Soukoulis C~M 2004 {\em Physical Review
  Letters\/} {\bf 93} 1--4

\bibitem{Simovski2007}
Simovski C~R 2007 {\em Metamaterials\/} {\bf 1} 62--80

\bibitem{Tsukerman2011}
Tsukerman I 2011 {\em Journal of the Optical Society of America B\/} {\bf 28}
  577

\bibitem{slovick2014generalized}
Slovick B~A, Yu Z~G and Krishnamurthy S 2014 {\em Physical Review B\/} {\bf 89}
  155118

\bibitem{Zhang2015a}
Zhang X and Wu Y 2015 {\em Scientific Reports\/} {\bf 5} 7892

\bibitem{Krokhin2016}
Krokhin A~A, Arriaga J, Gumen L~N and Drachev V~P 2016 {\em Physical Review
  B\/} {\bf 93} 75418

\bibitem{Tsukerman2017}
Tsukerman I 2017 {\em Physics Letters A\/} {\bf 381} 1635--1640

\bibitem{Liu2021SimultaneousMetasurfaces}
Liu Y, Ouyang C, Zheng P, Ma J, Xu Q, Su X, Li Y, Tian Z, Gu J, Liu L, Han J
  and Zhang W 2021 {\em ACS Applied Electronic Materials\/} {\bf 3} 4203--4209

\bibitem{Cui2019LightMetasurface}
Cui C, Yuan S, Qiu X, Zhu L, Wang Y, Li Y, Song J, Huang Q, Zeng C and Xia J
  2019 {\em Nanoscale\/} {\bf 11} 14446--14454

\bibitem{Ignatyeva2020All-dielectricResonances}
Ignatyeva D~O, Karki D, Voronov A~A, Kozhaev M~A, Krichevsky D~M, Chernov A~I,
  Levy M and Belotelov V~I 2020 {\em Nature Communications\/} {\bf 11} 5487

\bibitem{Abujetas2021Near-FieldModel}
Abujetas D~R and S{\'{a}}nchez-Gil J~A 2021 {\em Nanomaterials\/} {\bf 11} 998

\bibitem{Zhang2022HuygensCoating}
Zhang T, Duan Y, Huang L, Pang H, Liu J, Ma X, Shi Y and Lei H 2022 {\em
  Advanced Materials Interfaces\/} {\bf 9} 2102559

\bibitem{Brizi2022MagneticRegions}
Brizi D and Monorchio A 2022 {\em Scientific Reports\/} {\bf 12} 3258

\bibitem{Schilder2017}
Schilder N~J, Sauvan C, Sortais Y~R~P, Browaeys A and Greffet J~J 2017 {\em
  Physical Review A\/} {\bf 96} 13825

\bibitem{Pellico2019Nanoparticle-BasedImaging}
Pellico J, Ellis C~M and Davis J~J 2019 {\em Contrast Media {\&} Molecular
  Imaging\/} {\bf 2019} 1--13

\bibitem{Xie2018Shape-Theranostics}
Xie W, Guo Z, Gao F, Gao Q, Wang D, Liaw B~s, Cai Q, Sun X, Wang X and Zhao L
  2018 {\em Theranostics\/} {\bf 8} 3284--3307

\bibitem{Zhang2019ApplicationsCatalysis}
Zhang Q, Yang X and Guan J 2019 {\em ACS Applied Nano Materials\/} {\bf 2}
  4681--4697

\bibitem{Zhu2018MagneticApplications}
Zhu K, Ju Y, Xu J, Yang Z, Gao S and Hou Y 2018 {\em Accounts of Chemical
  Research\/} {\bf 51} 404--413

\bibitem{Thorat2017EffectiveRelease}
Thorat N~D, Bohara R~A, Noor M~R, Dhamecha D, Soulimane T and Tofail S~A~M 2017
  {\em ACS Biomaterials Science {\&} Engineering\/} {\bf 3} 1332--1340

\bibitem{Markel2012}
Markel V~A and Schotland J~C 2012 {\em Physical Review E\/} {\bf 85} 66603

\bibitem{Qi2019SynthesisActivity}
Qi H~P, Wang H~L, Zhao D~Y and Wang X~K 2019 {\em Materials Research
  Bulletin\/} {\bf 118} 110516

\bibitem{Du2020FeTherapy}
Du W, Liu T, Xue F, Cai X, Chen Q, Zheng Y and Chen H 2020 {\em ACS Applied
  Materials {\&} Interfaces\/} {\bf 12} 19285--19294

\bibitem{Guo20182DMedia}
Guo S, Xu F, Wang B, Wang N, Yang H, Dhanapal P, Xue F, Wang J and Li R~W 2018
  {\em Advanced Materials Interfaces\/} {\bf 5} 1800997

\bibitem{Mirabello2019UnderstandingMicroscopy}
Mirabello G, Keizer A, Bomans P~H~H, Kov{\'{a}}cs A, Dunin-Borkowski R~E,
  Sommerdijk N~A~J~M and Friedrich H 2019 {\em Chemistry of Materials\/} {\bf
  31} 7320--7328

\bibitem{Sihvola1990ChiralFormula}
Sihvola A and Lindell I 1990 {\em Electronics Letters\/} {\bf 26} 118

\bibitem{Lakhtakia1990DiluteSpheres}
Lakhtakia A, Varadan V~K and Varadan V~V 1990 {\em Applied Optics\/} {\bf 29}
  3627

\bibitem{Tretyakov1995MaxwellInclusions}
Tretyakov S and Mariotte F 1995 {\em Journal of Electromagnetic Waves and
  Applications\/} {\bf 9} 1011--1025

\bibitem{Weiglhofer1993Maxwell-garnettObjects}
Weiglhofer W~S, Lakhtakia A and Monzon J~C 1993 {\em Microwave and Optical
  Technology Letters\/} {\bf 6} 681--684

\bibitem{Sihvola1992AnalysisMixtures}
Sihvola A and Lindell I 1992 {\em Journal of Electromagnetic Waves and
  Applications\/} {\bf 6} 553--572

\bibitem{smith_analytic_2010}
Smith D~R 2010 {\em Physical Review E\/} {\bf 81} 036605

\bibitem{sozio_generalized_2015}
Sozio V, Vallecchi A, Albani M and Capolino F 2015 {\em Physical Review B\/}
  {\bf 91} 205127

\bibitem{alu_first-principles_2011}
Alu A 2011 {\em Physical Review B\/} {\bf 84} 075153

\bibitem{smith_composite_2000}
Smith D~R, Padilla W~J, Vier D~C, Nemat-Nasser S~C and Schultz S 2000 {\em
  Physical Review Letters\/} {\bf 84} 4184--4187

\bibitem{Bowen2017}
Bowen P~T, Baron A and Smith D~R 2017 {\em Physical Review A\/} {\bf 95} 33822

\bibitem{Wang2018AchievingParticles}
Wang B~X and Zhao C~Y 2018 {\em Physical Review A\/} {\bf 97} 023836

\bibitem{Ding1989EffectivePermittivities}
Ding K~H and Tsang L 1989 {\em Progress In Electromagnetics Research\/} {\bf
  01} 241--295

\bibitem{Shalaev1996ElectromagneticComposites}
Shalaev V~M 1996 {\em Physics Reports\/} {\bf 272} 61--137

\bibitem{deVries1998PointWaves}
de~Vries P, van Coevorden D~V and Lagendijk A 1998 {\em Reviews of Modern
  Physics\/} {\bf 70} 447--466

\bibitem{Barrera1988}
Barrera R~G, Monsivais G and Moch{\'{a}}n W~L 1988 {\em Physical Review B\/}
  {\bf 38} 5371--5379

\bibitem{chew1995waves}
Chew W~C 1995 {\em {Waves and fields in inhomogeneous media}\/} vol 522 (IEEE
  press New York)

\bibitem{mishchenko_multiple_2006}
Mishchenko M~I, Travis L~D and Lacis A~A 2006 {\em Multiple scattering of light
  by particles: radiative transfer and coherent backscattering\/} (Cambridge
  University Press)

\bibitem{Mackowski2013DirectParticles}
Mackowski D~W and Mishchenko M~I 2013 {\em Journal of Quantitative Spectroscopy
  and Radiative Transfer\/} {\bf 123} 103--112

\bibitem{ishimaru_theory_1977}
Ishimaru A 1977 {\em Proceedings of the IEEE\/} {\bf 65} 1030--1061

\bibitem{born2013principles}
Born M and Wolf E 2013 {\em {Principles of optics: electromagnetic theory of
  propagation, interference and diffraction of light}\/} (Elsevier)

\bibitem{Panasyuk2009}
Panasyuk G~Y, Schotland J~C and Markel V~A 2009 {\em Journal of Physics A:
  Mathematical and Theoretical\/} {\bf 42} 275203

\bibitem{Yaghjian1980}
Yaghjian A 1980 {\em Proceedings of the IEEE\/} {\bf 68} 248--263

\bibitem{sun2009novel}
Sun L~E and Chew W~C 2009 {\em Waves in Random and Complex Media\/} {\bf 19}
  162--180

\bibitem{Wang2021ClusteringParticles}
Wang L, Rasskazov I~L and Carney P~S 2021 {\em Physical Review B\/} {\bf 104}
  115418

\bibitem{Dolgaleva2012Local-fieldMaterials}
Dolgaleva K and Boyd R~W 2012 {\em Advances in Optics and Photonics\/} {\bf 4}
  1

\bibitem{Couture2010ExperimentalMetamaterial}
Couture S, Gauthier J, Kodera T and Caloz C 2010 {\em IEEE Antennas and
  Wireless Propagation Letters\/} {\bf 9} 1022--1025

\bibitem{Ozbay2008NegativeMetamaterial}
Ozbay E and Aydin K 2008 {\em Photonics and Nanostructures - Fundamentals and
  Applications\/} {\bf 6} 108--115

\bibitem{Gao2013MeasurementSample}
Gao P, Zhang C, Ai J, Li G and Kang Y 2013 {\em Physica A: Statistical
  Mechanics and its Applications\/} {\bf 392} 6506--6511

\bibitem{Chen2005NegativeMetamaterial}
Chen H, Ran L, Huangfu J, Zhang X, Chen K, Grzegorczyk T~M and Kong J~A 2005
  {\em Applied Physics Letters\/} {\bf 86} 151909

\bibitem{RAN2003Left-handedVerifications}
RAN L 2003 {\em Chinese Science Bulletin\/} {\bf 48} 1325

\bibitem{Aydin2005ObservationMetamaterials}
Aydin K, Guven K, Soukoulis C~M and Ozbay E 2005 {\em Applied Physics
  Letters\/} {\bf 86} 124102

\bibitem{Huangfu2004ExperimentalPatterns}
Huangfu J, Ran L, Chen H, Zhang X~m, Chen K, Grzegorczyk T~M and Kong J~A 2004
  {\em Applied Physics Letters\/} {\bf 84} 1537--1539

\bibitem{Wang2010ExperimentalTerahertz}
Wang S, Garet F, Blary K, Lheurette E, Coutaz J~L and Lippens D 2010 {\em
  Applied Physics Letters\/} {\bf 97} 181902

\bibitem{Chen2006MetamaterialWaves}
Chen H, Ran L, Wang D, Huangfu J, Jiang Q and Kong J~A 2006 {\em Applied
  Physics Letters\/} {\bf 88} 031908

\bibitem{Beruete2009NegativeSlab}
Beruete M, Navarro-C{\'{i}}a M, Sorolla M and Campillo I 2009 {\em Physical
  Review B\/} {\bf 79} 195107

\bibitem{Xi2009ExperimentalMetamaterial}
Xi S, Chen H, Jiang T, Ran L, Huangfu J, Wu B~I, Kong J~A and Chen M 2009 {\em
  Physical Review Letters\/} {\bf 103} 194801

\bibitem{Ozbay2007MetamaterialsSimulations}
Ozbay E, Guven K and Aydin K 2007 {\em Journal of Optics A: Pure and Applied
  Optics\/} {\bf 9} S301--S307

\bibitem{Smith2004MetamaterialsIndex}
Smith D~R 2004 {\em Science\/} {\bf 305} 788--792

\bibitem{Dong2005NumericalMetamaterials}
Dong Z~G, Zhu S~N, Liu H, Zhu J and Cao W 2005 {\em Physical Review E\/} {\bf
  72} 016607

\bibitem{Shelby2001ExperimentalRefraction}
Shelby R~A 2001 {\em Science\/} {\bf 292} 77--79

\bibitem{Sun2010LowArrays}
Sun J, Kang L, Wang R, Liu L, Sun L and Zhou J 2010 {\em New Journal of
  Physics\/} {\bf 12} 083020

\bibitem{Imhof2009ExperimentalMetamaterial}
Imhof C and Zengerle R 2009 {\em Appl Phys A\/} {\bf 94} 45--49

\bibitem{Smith2004EnhancedMetamaterial}
Smith D~R, Rye P~M, Mock J~J, Vier D~C and Starr A~F 2004 {\em Physical Review
  Letters\/} {\bf 93} 137405

\bibitem{mccall_negative_2002}
McCall M~W, Lakhtakia A and Weiglhofer W~S 2002 {\em European Journal of
  Physics\/} {\bf 23} 353

\bibitem{Bohren1998}
Bohren C~F and Huffman D~R 1998 {\em {Absorption and Scattering of Light by
  Small Particles}\/} (Weinheim, Germany: Wiley-VCH Verlag GmbH)

\bibitem{schwerdtfeger_table_2015}
Schwerdtfeger P 2015 {\em Centre for Theoretical Chemistry and Physics, Massey
  University\/}

\bibitem{Pulido-Mancera2017PolarizabilityModelingb}
Pulido-Mancera L, Bowen P~T, Imani M~F, Kundtz N and Smith D 2017 {\em Physical
  Review B\/} {\bf 96} 235402

\bibitem{Rockstuhl2007TheApproach}
Rockstuhl C, Zentgraf T, Pshenay-Severin E, Petschulat J, Chipouline A, Kuhl J,
  Pertsch T, Giessen H and Lederer F 2007 {\em Optics Express\/} {\bf 15} 8871

\bibitem{Jylha2006ModelingSpheres}
Jylh{\"{a}} L, Kolmakov I, Maslovski S and Tretyakov S 2006 {\em Journal of
  Applied Physics\/} {\bf 99} 043102

\bibitem{Liberal2017TheTechnologies}
Liberal I and Engheta N 2017 {\em Science\/} {\bf 358} 1540--1541

\bibitem{Liberal2017PhotonicMedia}
Liberal I, Mahmoud A~M, Li Y, Edwards B and Engheta N 2017 {\em Science\/} {\bf
  355} 1058--1062

\bibitem{Hinamoto2020ColloidalMetafluids}
Hinamoto T, Hotta S, Sugimoto H and Fujii M 2020 {\em Nano Letters\/} {\bf 20}
  7737--7743

\bibitem{gehring_observation_2006}
Gehring G~M, Schweinsberg A, Barsi C, Kostinski N and Boyd R~W 2006 {\em
  Science\/} {\bf 312} 895--897

\bibitem{bigelow_superluminal_2003}
Bigelow M~S, Lepeshkin N~N and Boyd R~W 2003 {\em Science\/} {\bf 301} 200--202

\bibitem{wang_gain-assisted_2000}
Wang L~J, Kuzmich A and Dogariu A 2000 {\em Nature\/} {\bf 406} 277--279

\bibitem{reitz2008foundations}
Reitz J~R, Milford F~J and Christy R~W 2008 {\em {Foundations of
  electromagnetic theory}\/} (Addison-Wesley Publishing Company)

\bibitem{Alu2005PolarizabilitiesLayers}
Al{\`{u}} A and Engheta N 2005 {\em Journal of Applied Physics\/} {\bf 97}
  094310

\bibitem{Werdehausen2018DesignNanocomposites}
Werdehausen D, Staude I, Burger S, Petschulat J, Scharf T, Pertsch T and Decker
  M 2018 {\em Optical Materials Express\/} {\bf 8} 3456

\end{thebibliography}

\end{document}